\documentclass[
 reprint,
bibnotes,
 amsmath,amssymb,
 aps,
dvipdfmx,
onecolumn
]{revtex4-2}

\usepackage{graphicx}
\usepackage{dcolumn}
\usepackage{bm}
\usepackage{braket}
\usepackage{cases}
\usepackage{amssymb}
\usepackage{color}
\usepackage{url}

\begin{document}

\title{Efficient criteria of quantumness for a large system of qubits}
 
\author{Shohei Watabe$^{1,2}$, Michael Zach Serikow$^{3,4}$, Shiro Kawabata$^5$, and Alexandre Zagoskin$^{6}$}  
\affiliation{$^1$ Department of Physics, Faculty of Science Division I, Tokyo University of Science, Shinjuku, Tokyo 162-8601, Japan}
\affiliation{$^2$ Division of Nano-quantum Information Science and Technology, Research Institute for Science and Technology, Tokyo University of Science, Shinjuku, Tokyo 162-8601, Japan}
\affiliation{$^3$ Department of Physics, University of Notre Dame, Notre Dame, IN 46556, USA} 
\affiliation{$^4$ Notre Dame Institute for Advanced Study, University of Notre Dame, IN 46556, USA} 
\affiliation{$^5$ Research Center for Emerging Computing Technologies (RCECT), National Institute of Advanced Industrial Science and Technology (AIST),
1-1-1 Umezono, Tsukuba, Ibaraki 305-8568, Japan.}
\affiliation{$^6$ Department of Physics, Loughborough University, Loughborough LE11 3TU, UK}

\begin{abstract} 
In order to model and evaluate large-scale quantum systems, {\it e.g.} quantum computer and quantum annealer, it is necessary to quantify the ``quantumness" of such systems. 
In this paper, we discuss the dimensionless combinations of basic parameters of large, partially quantum coherent systems, which could be used to characterize their degree of quantumness. Based on analytical and numerical calculations, we suggest one such number for a system of qubits undergoing adiabatic evolution, i.e., the {\it accessibility index}. Applying it to the case of D-Wave One superconducting quantum annealing device, we find that its operation as described falls well within the quantum domain.
\end{abstract}
\maketitle


\section{Introduction}

One of the key obstacles in the way to the full development of quantum technologies 2.0~\cite{Zagoskin2019} is the same circumstance which stimulated their development in the first place: the fundamental impossibility of an efficient  simulation of a large enough, quantum coherent structure with classical means. In practice ``large enough" turned out to be a system comprising a hundred or so quantum bits, which is still too small to form a quantum computer capable of simulating other ``large enough" quantum systems. On the other hand, artificial quantum coherent systems comprising thousands of qubits are being fabricated~\cite{Kakuyanagi2016} and even successfully used, like commercial quantum annealers~\cite{DWave2000X,Bunyk2014}. This ``quantum capacity gap"~\cite{Zagoskin2014b} needs to be bridged, in order to allow a systematic progress towards the development of the full potential of quantum technologies 2.0, such as noisy intermediate-scale quantum (NISQ) devices~\cite{Preskill2018quantumcomputingin} and universal fault tolerant quantum computers.  

The impossibility of an efficient classical simulation of a large quantum system is not absolute, in the sense that it concerns the simulation of an arbitrary evolution of such a system, whereby its state vector can reach all of its (exponentially high-dimensional) Hilbert space and has potentially infinite time to do so. The Margolus-Levitin theorem and its generalizations~\cite{Mandelstam1945,Margolus1998,Zwierz2012,Deffner2013,Deffner2013prl,Deffner2017} put a limit on the speed of such evolution, thus restricting the accessible part of the Hilbert space for any finite time interval. This agrees with a proof~\cite{Poulin2011} that the manifold of all quantum many-body states that can be generated by arbitrary time-dependent local Hamiltonians in a time that scales polynomially in the system size occupies an exponentially small volume in its Hilbert space. (This is a literally correct statement, since the  Hilbert space of a system of qubits is a finite-dimensional complex projective space; that is, it is compact and, moreover, it has a unitary invariant Fubini-Study metric~\cite{Brody2001}.) Numerical and analytical studies also indicate that the number of independent constraints describing quantum evolution may be much less than the dimensionality of the Hilbert space~\cite{Rigol2008}. It is therefore reasonable to suggest that a ``general case" evolution of a large quantum coherent quantum structure can be characterized by a non-exponentially large set of dimensionless parameters, which correspond to qualitatively different regimes of evolution of this structure. Our recent numerical simulations indicate the existence of such regimes in a set of qubits with pumping and dissipation~\cite{Andreev2019}. 

Such dimensionless parameters, if exist, will be combinations of fundamental physical constants and parameters, which characterize the system. (We will only need the Planck constant, since the speed of light and gravity constant are not relevant for the currently feasible devices). For example, in the standard approximation, a system of qubits is described by a quantum Ising Hamiltonian,  
\begin{align}
\hat{H}(t) = \sum_{i\neq j} J_{ij}(t) \sigma_i^z\sigma_j^z - \frac{1}{2}\sum_j \left(h_j(t) \sigma_j^z + \Delta_j(t) \sigma_j^x\right),
\label{eq:Ising1}
\end{align}
and a set of Lindblad operators responsible for dephasing and relaxation of separate qubits, with characteristic times, respectively, $t_{\phi}$ and $t_r$, and their combination, the decoherence time $t_D$. In the case of an adiabatic quantum processor~\cite{DWave2000X,Bunyk2014}, the time dependence of the Hamiltonian parameters $J_{ij}, h_j$ and $\Delta_j$ (except that induced by the ambient noise)  is determined by that of the adiabatic parameter $\lambda(t)$ (in the simplest case, by the time of adiabatic evolution, $t_f$). Then, the dimensionless characteristics of the system should be the combinations of $\hbar$ with the following quantities:
\begin{enumerate}
	\item Dimensionless:  $N$ (number of qubits); $\langle z\rangle$ (connectivity of the network: average number of couplings per qubit); $\langle\delta z^2\rangle$ (its dispersion); $\langle z_i z_j\rangle$ (its correlation function); $\dots$; 
	\item Powers of energy: $\langle E\rangle$ (average qubit excitation energy); $\langle \delta E^2\rangle$; $\langle E_i E_j\rangle$; $\langle J\rangle$ (average coupling strength); $\langle \delta J^2\rangle$; $\langle J_i J_k\rangle$; $k_BT$; $\dots$;
	\item Powers of time: $t_{\phi}$; $t_r$; $t_D$; $t_f$; $\dot{\lambda}  \sim 1/t_f$ (speed of evolution); $\ddot{\lambda}$; $\dots$;
	 \end{enumerate}
	and such additional parameters as, e.g., the spectral density of ambient noise $S_A(f)$. Note that all these parameters can be efficiently obtained by either direct measurements or straightforward calculations. 
	
The field of several dozens of  independent dimensionless combinations of the above parameters is narrowed  for a particular quantum system and the mode of its operation. Here we concentrate on adiabatic quantum computing (quantum annealing).  The Hamiltonian of a quantum system is here manipulated in such a way that its ground state changes from the easily accessible one to the one encoding a solution to the desired problem (and presumably having a very complex structure, so that reaching it by annealing would be improbable). In the case of a slow enough evolution of the Hamiltonian $H(t)$, the system initialized in the ground state of $H(0)$ will evolve into the ground state of $H(t_f)$ by virtue of the adiabatic theorem~\cite{Albash2018}. If the system is totally insulated, the quantum speed and accessibility theorems~\cite{Mandelstam1945,Margolus1998,Zwierz2012,Deffner2013,Deffner2013prl,Deffner2017,Poulin2011} do not put fundamental constraints on an adiabatic quantum computer. Nevertheless, in a realistic case, the computation time is limited by interactions with the outside world leading to nonunitarity~\cite{Sarandy2005a}, and the question arises whether the evolution from the initial to the desired final state of the system is possible. As we see, the question of accessibility of different regions of the Hilbert space is especially relevant in this case.

An intriguing twist is added by the fact that the operation of D-Wave processors demonstrated what looked convincingly like quantum annealing~\cite{Boixo2014,Albash2015} despite the large discrepancy between the adiabatic evolution time, $t_f$ (microseconds~\cite{Boixo2014}), and the qubit decoherence time, $t_D \ll t_f$ (tens of nanoseconds,~\cite{Boixo2013}), in the absence of any quantum error correction. On the second thought, it is not so surprising. The quantum state of an adiabatic quantum computer evolves starting at $t=0$ from a factorized state, $|{\rm in}\rangle$. The computation is successful, if at the time $t=t_f$ there is a sufficient ratio of quantum trajectories ending in the state $|{\rm out}\rangle$, which is also factorized by design. Decoherence tends to disrupt quantum correlations between different qubits, and thus constrain the trajectories to partially factorized submanifolds of the Hilbert space. Nevertheless the success does not necessarily require that these trajectories pass through globally entangled states, and thus certain degree of decoherence may not necessarily make the proper operation of the device impossible.  

While it cannot be predicted whether the evolution of a given quantum system can take it from the given initial to the desired final state,  the average of the maximal distance between \emph{some} initial and \emph{some} final state of the system, for given values of $t_f$ and other system parameters, may serve as a heuristic indicator of success. This distance can be naturally determined via the Fubini-Study metric~\cite{Brody2001}, in which the  distance $s(\phi,\psi)$ between states $|\phi\rangle$ and $|\psi\rangle$ is given by 
\begin{align}
\cos s(\phi,\psi) = \sqrt{\frac{\langle\phi|\psi\rangle \langle\psi|\phi\rangle}{\langle\phi|\phi\rangle\langle\psi|\psi\rangle} }.
\label{eq:FS}
\end{align}
The maximal Fubini-Study distance in the Hilbert space is $\pi/2$,  the distance between mutually orthogonal states. We will therefore choose the quantity 
\begin{align}
{\cal S} = \frac{2}{\pi}\sqrt{\overline{s^2(\phi,\psi)}}  
\label{eq:CONDITION 1}
\end{align}
as an \emph{ad hoc} parameter, which characterizes the ability of an adiabatically evolving quantum device to reach its desired quantum state. The bar denotes the averaging over all initial states and over all quantum trajectories accessible to the system, which connect them to the final states maximally removed from them (in terms of the Fubini-Study distance).

\section{Random walk model: Heuristic treatment}

The evolution of the state vector of a quantum system can be modeled by a series of  random collapses to one of its instantaneous eigenstates at the moments $t_1, t_2, \dots$, and unitary evolutions under the Hamiltonian $\hat{H}(t)$ between these moments. In the limit of infinitesimally small time intervals between subsequent collapses, this picture leads to the Quantum State Diffusion approach (QSD). Averaging over  individual QSD trajectories reproduces the standard quantum master equation for the density matrix and provides an efficient basis for numerical calculation~\cite{Percival1998}.

For our purpose it is essential to keep the decoherence time as an explicit parameter. Therefore, the model we use is a random walk in the Hilbert space of the system, with the step (Fubini-Study) lengths $\Delta s_j$ dependent on $\Delta t$. For small time intervals of unitary evolution $|\phi\rangle = e^{-i\hat{H}\Delta t}|\psi\rangle$, 
the distance $s(\phi,\psi)$ is given by (see Appendix~\ref{AppendixA})
\begin{align}
s \approx \frac{\Delta t}{\hbar} \sigma_{\hat H} (\psi). 
\label{eq:6}
\end{align} 
Here, $\sigma_{\hat H}^2(\psi) \equiv \langle \hat H^2\rangle_{\psi} - \langle \hat H\rangle_{\psi}^2$ is the energy dispersion during the unitary evolution, where $\langle \, \cdot \,  \rangle_\psi$ is the expectation value for a quantum state $| \psi \rangle $, i.e., $\langle \, \cdot \,  \rangle_\psi \equiv \langle \psi | \cdot | \psi \rangle$.  This expression is intuitively plausible: energy and time are the only parameters in the problem, and in case of zero energy variance the system would be in an eigenstate of the Hamiltonian and  remain in it, barring degeneracies. 
The energy variance is made possible by the interaction with the environment, which is implicit in the assumption of random collapses of the quantum state of the system. 

Substituting here the maximal possible value of $s = \pi/2$, we find 
\begin{align}
\Delta t_{\rm max} = \frac{\pi}{2} \frac{\hbar}{\sigma_{\hat H}(\psi)}.
\label{eq:7}
\end{align}
This coincides with the rigorous Mandelstam-Tamm expression for the minimal time  necessary to evolve from an initial state to a state orthogonal to it \cite{Mandelstam1945}.

Now consider the random walk of $M\gg 1$ steps of identical duration $\Delta t$, controlled by independently distributed random Hamiltonians $\hat H_j$ ($j=1,2,\dots M$) (but still assuming that the total displacement remains small). Using the same approximation as before, the span of this random walk (i.e., the Fubini-Study distance between its initial point $| \psi \rangle$ and final point $|\phi\rangle = \prod_j \exp{ (-i\Delta t  \hat H_j / \hbar )} |\psi\rangle$) is thus found directly (see Appendix~\ref{AppendixB}):
\begin{align}
s(\phi,\psi) \approx \frac{\Delta t}{\hbar} \sigma_{\sum_j \hat H_j}(\psi), 
\label{eq:13}
\end{align}
where 
\begin{align}
\sigma_{\sum_j \hat H_j}(\psi) = & 
\biggl \langle \sum_j \hat H_j^2 \biggr \rangle_\psi - \biggl \langle \sum_j \hat H_j \biggr \rangle_\psi^2
\\ 
= & \sum_{\chi\neq\psi} \sum_{jk} \langle\psi | \hat H_j|\chi\rangle\langle\chi|\hat H_k |\psi \rangle.
\end{align}
Here the summation is taken over the states $|\chi\rangle$ from the orthonormal basis of the Hilbert space of our systems, which includes the state $|\psi\rangle$.  

Making a further simplification, assume that all random steps have the same length $\Delta S = \overline{\sigma_{\hat H}} \Delta t/\hbar$. Then the r.m.s. of the span is
\begin{align}
\sqrt{\overline{s(\phi,\psi)^2}} \approx \Delta S \frac{\sqrt{\overline{\sigma_{\sum_j \hat H_j}^2(\psi)}}}{\overline{\sigma_{\hat H} (\psi)}},
\label{eq:14}
\end{align}
where the bar average is over random choice of $H_j$-Hamiltonians and initial states $|\psi\rangle$.

After averaging over random Hamiltonians $\hat H_j$ and states $|\psi\rangle$, which gives $\overline{\sum_j\langle\psi| \hat H_j|\chi\rangle} \approx 0$,   the only term surviving will be
$\overline{\sigma_{\sum \hat H_j}^2(\psi)} =\sum_{\chi\neq\psi}\sum_j  \overline{|\langle\psi| \hat H_j|\chi\rangle|^2} $. 
We thus have 
\begin{align}
\overline{\sigma_{\sum \hat H_j}^2(\psi)}  = & \sum_{j=1}^M \sum_{\chi\neq\psi} \overline{ |\langle\psi| \hat H_j|\chi\rangle|^2}
\\
\approx & M f(N) \overline{\sigma_{\hat H}(\psi)}^2. 
\label{eq:16}
\end{align}
Here, 
$f(N)$ is some function of the dimension of the Hilbert space $D=2^N$, which is given by 
$\sum_{\chi\neq\psi} \overline{ |\langle\psi| \hat H_j|\chi\rangle|^2}
\equiv f(N) \overline{\sigma_{\hat H} (\Psi)}^2$, where we used the fact that the averaged value with respect to the random $\hat H_j$ is $j$-independent. 

Then we obtain from Eq. (\ref{eq:CONDITION 1})  
\begin{align}
{\cal S} \approx \frac{2}{\pi}\Delta S M^{1/2} f(N)^{1/2}.
\label{eq:17}
\end{align}
In particular, the condition ${\cal S}=1$ yields the relation between $M$ and $\Delta S$, for which the random walk is likely to connect mutually orthogonal states in the Hilbert space (a \emph{maximal random walk}), and thus the quantum adiabatic operation of the system we model should be possible. This ``critical value" of $\Delta S$ for a given $M, N$ is
\begin{align}
\Delta S = \frac{\pi}{2} M^{-1/2} f(N)^{-1/2}.
\label{eq:critical}
\end{align}

\begin{figure*}[tbp]
      \centering
      \includegraphics[width=160mm]{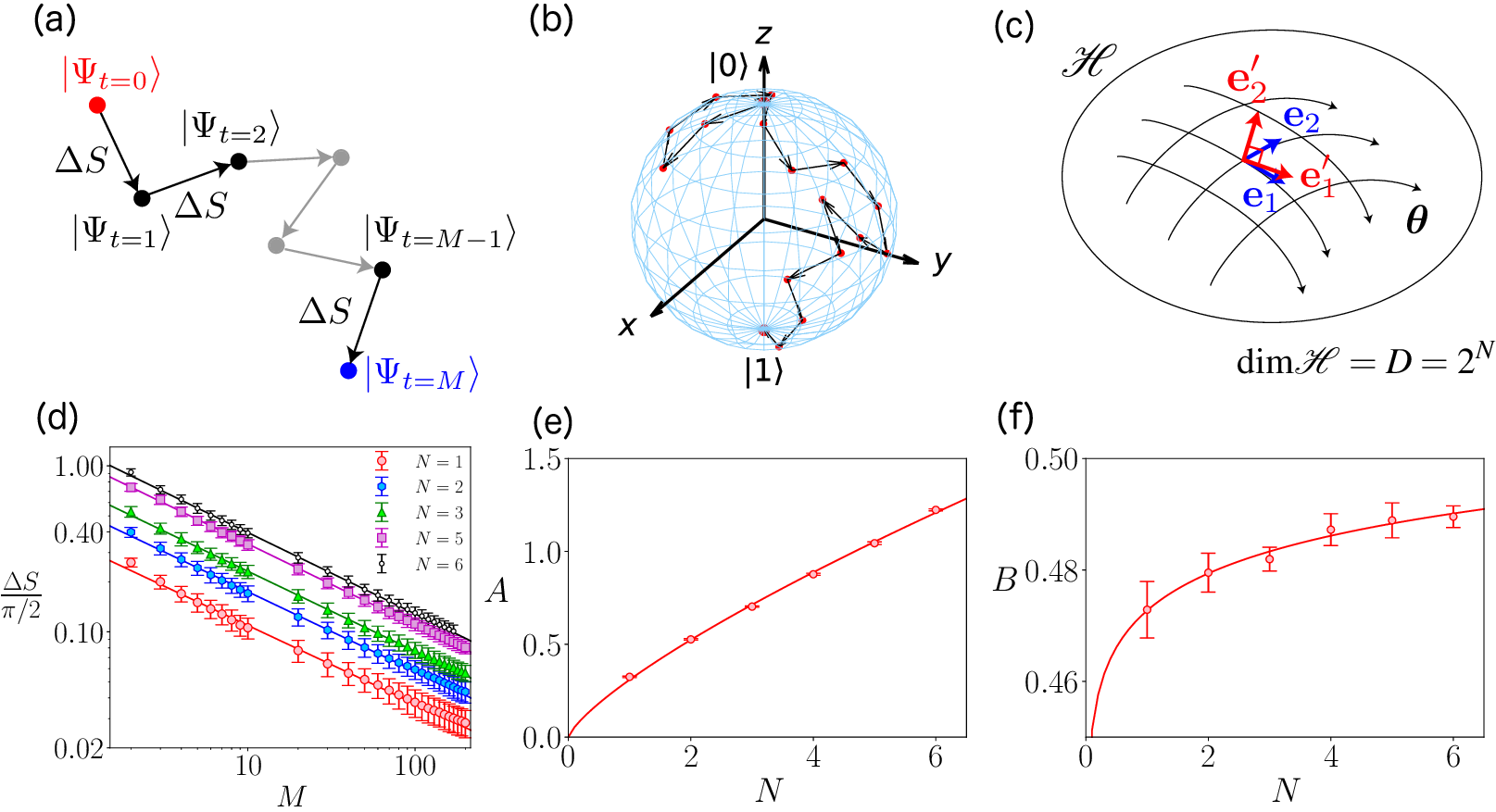}
      \caption{
      Random walk in the Hilbert space. 
      (a) Schematics of states walking randomly in the Hilbert space. The points represent states. The state walks randomly from $|\Psi_{t=0}\rangle$ to $| \Psi_{t=M}\rangle$, where the stride of each step is given by the Fubini-Study distance $\Delta S$. 
      (b) Schematic of the random walk in the Hilbert space in the Bloch sphere. 
      The red points indicate the states, which walks with the random direction with a fixed stride $\Delta S$. 
      The random walker on the Bloch sphere starts from the initial state $|0\rangle $ at the north pole, 
      and finally reaches the state $|1 \rangle$ at the south pole. 
      (c) For considering the uniform random walk in the Hilbert space, we generate a random vector in the local orthogonal coordinate (spanned by the bases ${\bf e}_{1,2}'$, where ${\bf e}_{1}'\cdot{\bf e}_{2}' = 0$, for example) from the curved parameter space ${\bm \theta}$ (spanned by the bases ${\bf e}_{1,2}$, where ${\bf e}_{1}\cdot{\bf e}_{2} \neq 0$). 
      (d) The critical Fubini-Study distance $\Delta S$ for the random walk in the in the Hilbert space, as a function of the number of random steps $M$. 
      We consider $N$-qubits states. 
      The data points are the averaged value for $500$ trials.
      Solid lines are fitted lines with $\Delta S = (\pi/2)A M^{-B}$. 
The fitting is performed for $M \geq 3$. 
       (e) The fitting factor $A$ and (f) the fitting factor $B$ as a function of the number of qubits $N$. 
      The solid lines represent fitting functions, where we used $A = \alpha_A N^{\beta_A}$ and $B = \alpha_B N^{\beta_B}$ for $N=1$--$6$. 
      }
      \label{Fig_randomwalk}
\end{figure*}

\section{Random walk model: Numerical approach}

The goal of our numerical simulations is to determine the function $f(N)$ in Eq. \eqref{eq:critical} through the relation between the Fubini-Study length of a single step, $\Delta S$, and the number of steps, $M$, of a maximal random walk in the $D=2^N$-dimensional Hilbert space (see Figs.~\ref{Fig_randomwalk} (a) and (b)).

A $D$-dimensional quantum state can be parameterized by $2(D-1)$ real parameters ${\bm \theta} = (\theta_1, \theta_2, \cdots, \theta_{2 (D-1)})^{\rm T}$, because of the presence of the overall phase factor and the normalized condition. 
The state can be given by 
		\begin{align}
		| \Psi ( {\bm \theta})\rangle 
		=  
		\left ( \begin{array}{rl}
		e^{i\theta_D} &\cos(\theta _{1})
		\\
		e^{i\theta_{D+1}} &\sin(\theta _{1})\cos(\theta _{2})
		\\
		e^{i\theta_{D+2}} &\sin(\theta _{1})\sin(\theta _{2})\cos(\theta _{3})
		\\
		 & \vdots
		 \\
		e^{i\theta_{2(D-1)}} & \sin(\theta _{1})\cdots \sin(\theta _{D-2})\cos(\theta _{D-1})
		\\ 
		& \sin(\theta _{1})\cdots \sin(\theta_{D-2})\sin(\theta _{D-1})
		\end{array}
		\right ). 
		\label{eqB1}
		\end{align}

We consider the uniform random walk in the Hilbert space using the form \eqref{eqB1}, where each step length is fixed with the Fubini-Study distance $\Delta S$. 
First, we randomly prepare an initial state $|\Psi_{t=0} \rangle$ in the $D$-dimensional Hilbert space. 
The state $|\Psi_t ({\bm \theta}) \rangle$ at a step $t$ is parameterized by ${\bm \theta}$. The state at $t+1$ is updated as $|\Psi_{t + 1}({\bm \theta} + d {\bm \theta}) \rangle $. 

The Fubini-Study distance between these states $| \Psi ({\bm \theta} + d {\bm \theta}) \rangle $ and $| \Psi ({\bm \theta}) \rangle $ can be written as 
\begin{align}
\Delta S = s(\Psi_t ({\bm \theta}),\Psi_{t + 1}({\bm \theta} + d {\bm \theta}))  =\sum_{ij}  g_{ij}d\theta_i d \theta_j, 
\label{eq:next2}
\end{align}
where the Fubini-Study metric is given by 
\begin{align}
	g_{ij} ( {\bm \theta}) = & \langle \partial_{\theta_i} \Psi  ( {\bm \theta}) | \partial_{\theta_j} \Psi  ( {\bm \theta}) \rangle 
	\nonumber 
	\\ 
	& - \langle \partial_{\theta_i} \Psi  ( {\bm \theta})| \Psi  ( {\bm \theta})\rangle \langle \Psi  ( {\bm \theta})| \partial_{\theta_j} \Psi  ( {\bm \theta})\rangle, 
	\label{eqBB16}
\end{align}
with $| \partial_{\theta_i} \Psi ({\bm \theta}) \rangle \equiv \partial | \Psi ({\bm \theta}) \rangle / \partial \theta_i$ for $i,j = 1, 2, \cdots, 2 (D-1)$. 
We generate the parameter $d {\bm \theta}$ such that the Fubini-Study distance $\sum_{ij} g_{ij}d\theta_i d \theta_j$ satisfies the given fixed value $\Delta S$, and 
the direction of the random step is uniformly random in the local orthogonal space, where the parameters ${\bm \theta}$ are given in the non-orthogonal coordinates with the curved space
(See Figs.~\ref{Fig_randomwalk} (c) and Appendix~\ref{AppendixRandomWalk}). 
Since we use the form \eqref{eqB1}, 
the state vector remains normalized. 

The span of the random walk from $t=0$ to $M$ is then given by 
\begin{align}
s(\Psi_0,\Psi_M) = \arccos(\langle\Psi_0|\Psi_M\rangle).
\label{eq:next3}
\end{align}
In our simulations we fix the number of steps $M$ and search for such a value of $\Delta S$ (``critial value") that the final state $| \Psi_{t=M} \rangle$  satisfies the condition $| \pi/2 - s(\Psi_0,\Psi_M) | \leq \epsilon$ for some small $\epsilon$. 

Figure~\ref{Fig_randomwalk} (d) shows the critical value of $\Delta S$ as a function of $M$. Fitting yields
\begin{align}
\Delta S = \frac{\pi}{2} A(N) M^{-B(N)}, 
\label{eq:next4}
\end{align}
where $A(N) =0.309(8) N^{0.76(2)}$ and $B(N) = 0.4727(8) N^{0.020(1)}$ (see Figs~\ref{Fig_randomwalk} (e) and (f)). In these numerical simulations, we used $\epsilon = \pi/2 - \cos^{-1} (1/\sqrt{D})$, where $\epsilon \approx 1/\sqrt{D}$ in the huge Hilbert space.  We see that the critical value of $\Delta S$ scales almost as a power of $M$, since $B(N)$ shows a very weak power law dependence on the number of qubits. 

Comparing Eqs.(\ref{eq:next4}) and (\ref{eq:critical}), we see that the dependence on $M$ in our heuristic and numerical approaches is almost the same, while for the function $f(N)$ we find $f(N)^{} \approx 10.5 \:N^{-3/2}$.

\section{Accessibility index, "quantumness" criterion and comparison to experiment}

The quantities $\Delta S$ and $M$ do not have a direct experimental significance. In our approximate treatment we can relate them instead to the decoherence time (during which the unitary evolution takes place), $t_D$, and the total time of adiabatic evolution, $t_f$, via
\begin{align}
M = \frac{t_f}{t_D}; \:\: \Delta S \leq \frac{t_D}{\hbar} J,
\label{eq:next5}
\end{align}
where $J$ is the typical coupling between qubits, so that $J$ is a reasonable measure of the uncertainty of the $N$-qubit system's energy during the adiabatic evolution (here we use the Mandelstam-Tamm expression). Then from (\ref{eq:next4}) we find, that the necessary ``quantumness" criterion for the adiabatic evolution is
\begin{align}
t_D\frac{2\pi J}{h}  >  \frac{\pi}{2} C N^{3/4} \left(\frac{t_D}{t_f}\right)^{1/2},
\label{eq:next6}
\end{align}
where $C \approx 0.31$. Given all the approximations we have made, we can as well take $C = 1$. Then the "quantumness" condition can be written as 
\begin{equation}
{\cal A} > 1,
\label{eq:next7a}
\end{equation}
where the \emph{accessibility index} for a system of $N$ qubits with average coupling strength $J$ is
\begin{equation}
{\cal A} \equiv   4\frac{J}{h}\frac{\sqrt{t_f t_D} }{ N^{3/4}}. 
\label{eq:}
\end{equation}

Applying this criterion to the operation of the D-Wave processor described in Refs.~\cite{Boixo2013,Boixo2014}, with $t_D \sim 10$ ns, $t_f\sim 5\mu$s, $N \sim 100$ and $J/h \sim 5$GHz, we see that 
\begin{align}
{\cal A} \approx 10^2 \gg 1,
\label{eq:next8}
\end{align}
and the necessary ``quantumness" condition was satisfied. This indicates that the results of Refs.~\cite{Boixo2013,Boixo2014} are consistent with quantum annealing. From Eq. \eqref{eq:next7a}, we can evaluate the maximal size of a quantum processor for which the "quantumness" condition holds, other things being the same as in ~\cite{Boixo2013,Boixo2014}: $N_{\rm max} \equiv  [4 J/h]^{4/3} (t_f t_D)^{2/3}\approx 7 \times 10^4$.

Note that the condition ${\cal A}>1$ does not guarantee the successful operation of a quantum processor, i.e., it reaching the desired final state. It is rather the necessary, but not sufficient, condition of success.


\section{Conclusions}

We have investigated the generic behavior of a partially coherent system of qubits undergoing adiabatic evolution. Our aim was finding a convenient dimensionless parameter, which could characterize the degree of ``quantumness" of our system.  
Basing our analysis on a random walk model of quantum evolution in the Hilbert space, we found a parameter allowing to evaluate the likelihood of a successful quantum transition between the initial and desired final states of the system. It is the \emph{accessibility index}, expressed through the qubit decoherence time, time of evolution and the qubit coupling strength. Applying it to the case of 128-qubit D-Wave processors, we found that their evolution was consistent with quantum adiabatic transitions despite the qubit decoherence time being much smaller than the evolution time. An important future study is to assess the effectiveness of the \emph{accessibility index} in more detail by considering a specific Hamiltonian system coupled to an environment.

\begin{acknowledgments}
This paper is partly based on results obtained from a project, JPNP16007, commissioned by the New Energy and Industrial Technology Development Organization (NEDO), Japan. S.W. was supported by Nanotech CUPAL, National Institute of Advanced Industrial Science and Technology (AIST). A.Z. was supported by NDIAS Residential Fellowship. 
\end{acknowledgments}


\bibliography{draft.bib}

\appendix

\section{Derivation of Eq.~\eqref{eq:6}}\label{AppendixA}

In this appendix, we derive Eq.~\eqref{eq:6}. 
From Eq.~\eqref{eq:FS}, 
the cosine distance between $| \phi \rangle$ and $|\psi \rangle$ is given by 
$ \cos s(\phi,\psi) = \sqrt{\langle\phi|\psi\rangle \langle\psi|\phi\rangle/ (\langle\phi|\phi\rangle\langle\psi|\psi\rangle)}$. 
For small time intervals of unitary evolution, we have 
\begin{align}
|\phi\rangle = & e^{-i\hat H \Delta t / \hbar}|\psi\rangle 
\\ 
= & \left [1-\frac{i}{\hbar} \hat H\Delta t - \frac{1}{2\hbar^2} \hat H^2(\Delta t)^2 \right ] |\psi\rangle + o((\Delta t)^2). 
\label{eq:FSsmall} 
\end{align} 
Here, 
\begin{align}
\langle\phi|\psi\rangle \langle\psi|\phi\rangle 
= & \left|\langle\psi| 
 \left [1-\frac{i}{\hbar} \hat H\Delta t - \frac{1}{2\hbar^2} \hat H^2(\Delta t)^2 \right ] |\psi\rangle\right|^2 + o((\Delta t)^2) 
\\ 
= & 
\left|1-\frac{i\Delta t}{\hbar}\langle \hat H\rangle_{\psi} - \frac{(\Delta t)^2}{2\hbar^2}\langle \hat H^2\rangle_{\psi}\right|^2 + o((\Delta t)^2)  
\\ 
= & 
1-\frac{(\Delta t)^2}{\hbar^2}\langle \hat H^2\rangle_{\psi} +\frac{(\Delta t)^2}{\hbar^2}\langle \hat H\rangle_{\psi}^2  + o((\Delta t)^2) 
\\ 
= & 
1 - \frac{(\Delta t)^2}{\hbar^2} \sigma_{\hat H}^2(\psi) + o((\Delta t)^2).
\label{eq:3}
\end{align}
where $\sigma_{\hat H}^2(\psi) \equiv \langle \hat H^2\rangle_{\psi} - \langle \hat H\rangle_{\psi}^2$. On the other hand, $\langle\psi|\psi\rangle = 1$ and 
\begin{align}
 \langle\phi|\phi\rangle = & 
 \langle\psi| 
 \left [ 1+\frac{i}{\hbar}\hat H\Delta t - \frac{1}{2\hbar^2} \hat H^2(\Delta t)^2 \right ] 
 \left [ 1-\frac{i}{\hbar}\hat H\Delta t - \frac{1}{2\hbar^2} \hat H^2(\Delta t)^2 \right ] 
 |\psi\rangle + o((\Delta t)^2)
 \\ 
= & 
 \langle\psi| \left [ 1+\frac{1}{\hbar^2}\hat H^2(\Delta t)^2 - 2\frac{1}{2\hbar^2}\hat H^2(\Delta t)^2 \right ] |\psi\rangle + o((\Delta t)^2) 
 \\ 
= & 
 1 + o ((\Delta t)^2).
\label{eq:4}
\end{align}
We can write the right hand side of Eq.~\eqref{eq:FS} as 
\begin{align}
\sqrt{\frac{\langle\phi|\psi\rangle \langle\psi|\phi\rangle}{\langle\phi|\phi\rangle\langle\psi|\psi\rangle} }
= & 
\sqrt{1 - \frac{(\Delta t)^2}{\hbar^2} \sigma_H^2(\psi)} +o((\Delta t)^2) 
\\ 
= & 
1 - \frac{(\Delta t)^2}{2\hbar^2} \sigma_H^2(\psi) +o((\Delta t)^2),
\label{eq:AppendixA5}
\end{align}
Since $s = O(\Delta t)$, 
we have 
\begin{align}
\cos s(\phi,\psi) = 1 - \frac{1}{2}s^2 +o((\Delta t)^2) . 
\label{eq:AppendixA6}
\end{align}
Therefore using Eqs~\eqref{eq:FS}, \eqref{eq:AppendixA5}, and \eqref{eq:AppendixA6}, 
we can obtain Eq.~\eqref{eq:6}. 

\section{Derivation of Eq.~\eqref{eq:13}}\label{AppendixB}

Using the same approximation as in Appendix~\ref{AppendixA}, we obtain 
\begin{align}
|\phi\rangle = & \prod_j \exp\left(\frac{-i\Delta t}{\hbar} \hat H_j  \right)|\psi\rangle 
\\ 
\approx & \prod_j \left(1-\frac{i\Delta t}{\hbar} \hat H_j - \frac{(\Delta t)^2}{2\hbar^2}  \hat  H_j^2  \right)|\psi\rangle
\\ 
\approx & 
\left(1  - \frac{i\Delta t}{\hbar} \sum_j \hat H_j - \frac{(\Delta t)^2}{2\hbar^2} \sum_j \hat H_j^2 - \frac{(\Delta t)^2}{\hbar^2} \sum_{j>k} \hat H_j \hat H_k \right)|\psi\rangle. 
\label{eq:8}
\end{align} 
Here, we have 
\begin{align}
|\langle\psi|\phi\rangle|^2 
\approx & 
\left|\left(1 - \frac{i\Delta t}{\hbar} \sum_j \langle\psi| \hat H_j|\psi\rangle   - \frac{(\Delta t)^2}{2\hbar^2} \sum_j \langle\psi| \hat H_j^2|\psi\rangle   - \frac{(\Delta t)^2}{\hbar^2} \sum_{j>k} \langle\psi| \hat  H_j \hat H_k|\psi\rangle    \right)\right|^2 
\\ 
= & 
\left(1- \frac{(\Delta t)^2}{2\hbar^2} \sum_j \langle\psi| \hat  H_j^2|\psi\rangle   - \frac{(\Delta t)^2}{\hbar^2} {\tt Re}\sum_{j>k} \langle\psi| \hat H_j\hat H_k|\psi\rangle    \right)^2 
\\ 
& + 
\left( \frac{\Delta t}{\hbar}\sum_j \langle\psi|\hat H_j|\psi\rangle  + \frac{(\Delta t)^2}{\hbar^2}{\tt Im}\sum_{j>k} \langle\psi|\hat H_j\hat H_k|\psi\rangle \right)^2.
\label{eq:9}
\end{align}
Since 
\begin{align}
& \frac{(\Delta t)^2}{2\hbar^2} \sum_j \langle\psi| \hat H_j^2|\psi\rangle  + \frac{(\Delta t)^2}{\hbar^2} {\tt Re}\sum_{j>k} \langle\psi| \hat H_j \hat H_k|\psi\rangle 
\\ 
= & \frac{(\Delta t)^2}{2\hbar^2} \sum_j \langle\psi| \hat H_j^2|\psi\rangle  + \frac{(\Delta t)^2}{2\hbar^2}\sum_{j\neq k} \langle\psi| \hat H_j\hat H_k|\psi\rangle 
\\  
= & 
\frac{(\Delta t)^2}{2\hbar^2}
\langle\psi | \biggl (\sum_{j} \hat H_j \biggr )^2 | \psi \rangle, 
\label{eq:11}
\end{align}
we finally have 
\begin{align}
|\langle\psi|\phi\rangle|^2  = & 1 -  \frac{(\Delta t)^2}{\hbar^2}
\langle\psi | \biggl (\sum_{j} \hat H_j \biggr )^2 | \psi \rangle 
+ \frac{(\Delta t)^2}{\hbar^2} \biggl ( \langle\psi |  \sum_{j} \hat H_j   | \psi \rangle \biggr )^2 + o((\Delta t)^2) 
\\ 
= & 
1- \frac{(\Delta t)^2}{\hbar^2} \sigma_{\sum  \hat H_j}^2(\psi) + o((\Delta t)^2).
\label{eq:12}
\end{align}

The span of this random walk is thus given by 
\begin{align}
\cos s(\phi,\psi) = & |\langle\psi|\phi\rangle| + o((\Delta t)^2) 
\\ 
= & 1 - \frac{(\Delta t)^2}{2\hbar^2} \sigma_{\sum  \hat H_j}^2(\psi) + o((\Delta t)^2). 
\label{eq:B13}
\end{align}
By using Eqs.~\eqref{eq:AppendixA6} and \eqref{eq:B13}, we obtain Eq.~\eqref{eq:13}. 
Here, we also have the representation 
\begin{align}
\sigma_{\sum  \hat H_j}^2(\psi)= & 
\sum_{jk}\left\langle\psi\left|H_jH_k\right|\psi\right\rangle 
- \biggl ( \sum_j \left\langle\psi\left|H_j\right|\psi\right\rangle \biggr )^2 
\\ 
= & 
\sum_{\chi} \sum_{jk}\left\langle\psi\left|H_j|\chi\rangle\langle\chi|H_k\right|\psi\right\rangle - \biggl ( \sum_j \left\langle\psi\left|H_j\right|\psi\right\rangle \biggr ) \biggl ( \sum_{k} \left\langle\psi\left|H_k\right|\psi\right\rangle\biggr )  
\\ 
= & 
 \sum_{\chi\neq\psi} \sum_{jk}\left\langle\psi\left|H_j|\chi\rangle\langle\chi|H_k\right|\psi\right\rangle.
\label{eq:15}
\end{align}

\section{Algorithm for random walk in Hilbert space}\label{AppendixRandomWalk}

A $D$-dimensional quantum state can be parameterized by $2(D-1)$ real parameters ${\bm \theta} = (\theta_1, \theta_2, \cdots, \theta_{2 (D-1)})^{\rm T}$, 
where a state can be givne by \eqref{eqB1}. 
The distance between states $| \Psi ({\bm \theta} + d {\bm \theta}) \rangle $ and $| \Psi ({\bm \theta}) \rangle $ can be given by  
	\begin{align}
		\Delta S^2 = g_{ij} ( {\bm \theta}) d \theta_i d \theta_j =d {\bm \theta}^{\rm T} {\bf g} ( {\bm \theta}) d{\bm \theta}, 
	\end{align}
	where the Fubini-Study metric $[{\bf g}( {\bm \theta})]_{ij} = g_{ij}$ is given in Eq. \eqref{eqBB16} and $d {\bm \theta}$ is given by $d {\bm \theta} = (d \theta_1, d \theta_2, \cdots,d \theta_{2 (D-1)})^{\rm T}$. 
Using the form in Eq. (\ref{eqB1}), we find that $g_{ij} \in {\mathbb R}$ for $i,j \leq D-1$ and $i,j \geq D$. 
We also have $(g_{ij} + g_{ji}) d\theta_i d \theta_{j} =0$. We can thus take ${\bf g}$ as a real symmetric matrix by replacing $g_{ij}, g_{ji}$ with $(g_{ij} + g_{ji})/2$. 

We then consider the uniform random walk in the Hilbert space, 
where each step length is fixed with the Fubini-Study distance $\Delta S$. 
We first generate a random initial state $| \Psi_{t=0}  \rangle$ with the random phase and the random amplitude that satisfies the normalization condition. 
The parameters $\theta_{i = 1, \cdots, 2(D-1)}$ are obtained by inversely solving Eq.~\eqref{eqB1}.  
Since the parameters ${\bm \theta}$ are given in the non-orthogonal coordinates with the curved space, 
we consider the local coordinate transformation to the local orthogonal coordinate at the point ${\bm \theta}$ in order to consider the uniformly distributed random walk in the Hilbert space. 
In this case, the distance $\Delta S$ can be given by 
	\begin{align}
		\Delta S^2 
		= & \Delta {\bm \Theta}^{\rm T} {\bf h} \Delta {\bm \Theta}
		=  \sum_{i =1}^{2 (D-1)} h_i \Delta \Theta_i^2, 
		\label{eqB4}
	\end{align}
where we used the fact that the real symmetric matrix ${\bf g}$ is diagonalized by orthogonal matrix ${\bf V}$ as ${\bf V}^{\rm T} {\bf g} {\bf V} = {\bf h} = {\rm diag}[h_1, h_2, \cdots, h_{2(D-1)}]$ with $h_i \in {\mathbb R}$, 
${\bf g} {\bf v}_i = h_i {\bf v}_i$,  ${\bf V} = ({\bf v}_1, {\bf v}_2, \cdots, {\bf v}_{2(D-1)})$, and $\Delta {\bm \Theta} = {\bf V}^{\rm T} \Delta {\bm \theta}$. 
Since ${\bf v}_i \in {\mathbb R}^{2(D-1)}$, we also have $\Delta {\bm \Theta}  \in {\mathbb R}^{2(D-1)}$. 
We generate the vector ${\bf u} = (u_1, u_2, \cdots, u_{2(D-1)})^{\rm T}$ uniformly distributed over the surface of the unit $2(D-1)$-ball, which can easily constructed from the $2(D-1)$-real random values $(x_1, x_2, \cdots, x_{2(D-1)})$ generated from the normal distribution ${\mathcal N}(0, 1)$ with the relation $u_i = x_i / \sqrt{x_1^2 + x_2^2 + \cdots + x_{2(D-1)}^2}$. 
Using the uniformly distributed unit vector ${\bf u}$, we can take 
	\begin{align}
		\Delta \Theta_i = \frac{ \Delta S  }{ \sqrt{h_i} } u_i, 
	\end{align}
which satisfies the relation \eqref{eqB4}. 
The inverse transformation provides $\Delta {\bm \theta} = {\bf V} \Delta {\bm \Theta}$. 

By employing a state $| \Psi_t ({\bm \theta}) \rangle$, 
we can generate a new state $| \Psi_{t+1} ({\bm \theta} + \Delta {\bm \theta}) \rangle$ from one of the uniformly distributed random vectors ${\bf u}$, the distance of which between $| \Psi_t \rangle$ and $| \Psi_{t+1} \rangle$ is $\Delta S$. 
The Fubini-Study distance between the state $|\Psi_{t} \rangle $ and the initial state $|\Psi_{t=0}\rangle = |0 \rangle$ is given by 
$L_t = \cos^{-1} (| \langle \Psi _t | 0 \rangle|)$. 
At the random walk step $t=M$, if the state $| \Psi_{t=M} \rangle$ reaches the distance $\pi/2 - L_{t=M} \leq \epsilon$ with a small parameter, 
we consider that the state starting from $|\Psi_{t=0}\rangle = |0 \rangle$ arrives a farthest state through the random walk.

\end{document}